\documentclass[12pt]{article}

\usepackage{tocloft}
\usepackage{graphicx}
\graphicspath{ {./Plots/} }
\usepackage{natbib}
\usepackage[protrusion=true,expansion=true]{microtype}
\usepackage{amsmath,amsthm,amsfonts,amssymb,mathrsfs,dsfont,mathtools}
\usepackage{booktabs}
\usepackage{dutchcal}
\usepackage{url}
\usepackage{float}
\usepackage{color}
\usepackage{parskip}
\usepackage[explicit]{titlesec}
\usepackage{threeparttable}
\usepackage{setspace}
\usepackage[margin=0.85in]{geometry}
\usepackage[font = onehalfspacing]{caption}
\usepackage[usenames,dvipsnames]{xcolor}
\usepackage[hyperfootnotes=true]{hyperref}
\usepackage[labelfont=bf]{caption}
\usepackage{authblk}
\usepackage{caption}
\usepackage[titletoc,toc,title]{appendix}
\usepackage{enumerate}
\usepackage{soul}
\usepackage{subfig}
\usepackage{accents}
\usepackage{subfiles} 
\usepackage[capitalise, nameinlink]{cleveref}
\usepackage{dsfont}
\usepackage{comment}

\usepackage{changepage}
\usepackage[noend]{algpseudocode}
\usepackage{tocloft}
\usepackage[protrusion=true,expansion=true]{microtype}
\usepackage{amsmath,amsthm,amsfonts,amssymb,mathrsfs,dsfont,mathtools}
\usepackage{float}
\usepackage{color}
\usepackage{parskip}
\usepackage[explicit]{titlesec}
\usepackage{threeparttable}
\usepackage{setspace}
\usepackage[font = onehalfspacing]{caption}
\usepackage[usenames,dvipsnames]{xcolor}
\usepackage[hyperfootnotes=true]{hyperref}
\usepackage[labelfont=bf]{caption}
\usepackage{authblk}
\usepackage[titletoc,toc,title]{appendix}
\usepackage{enumerate}
\usepackage{soul}
\usepackage{subfig}
\usepackage{accents}
\usepackage[capitalise, nameinlink]{cleveref}
\usepackage{diagbox}
\usepackage{array,multirow}

\usepackage[utf8]{inputenc} 
\usepackage[T1]{fontenc}    
\usepackage{hyperref}       
\usepackage{url}            
\usepackage{booktabs}       
\usepackage{amsfonts}       
\usepackage{nicefrac}       
\usepackage{microtype}      
\usepackage{lipsum}		
\usepackage{graphicx}
\usepackage{natbib}
\usepackage{doi}
\usepackage{subfiles}
\usepackage{bbm}

\usepackage{adjustbox}
\usepackage{multirow}
\usepackage{amsmath}
\usepackage[section]{placeins}
\usepackage{algorithm}
\usepackage{algpseudocode}
\usepackage{dsfont}

\DeclareUnicodeCharacter{2212}{-}

\newcommand{\rc}[1]{{\color{red}{#1}}}

\setstcolor{red} 

\hypersetup{colorlinks = true, citecolor = MidnightBlue, linkcolor = MidnightBlue, urlcolor = MidnightBlue}

\titleformat{\section}{\normalfont\large\bfseries}{\thesection}{1em}{#1}
\titleformat{\subsection}{\normalfont\normalsize\bfseries}{\thesubsection}{1em}{#1}
\titleformat{\subsubsection}{\normalfont\normalsize\itshape}{\thesubsubsection}{1em}{#1}
\titlespacing\section{0pt}{12pt plus 4pt minus 2pt}{6pt plus 2pt minus 2pt}
\titlespacing\subsection{0pt}{12pt plus 4pt minus 2pt}{3pt plus 2pt minus 3pt}
\titlespacing\subsubsection{0pt}{12pt plus 4pt minus 2pt}{0pt plus 2pt minus 3pt}

\def\boxit#1{\vbox{\hrule\hbox{\vrule\kern6pt
          \vbox{\kern6pt#1\kern6pt}\kern6pt\vrule}\hrule}}

\definecolor{orange}{rgb}{1,0.5,0}
\definecolor{MyDarkBlue}{rgb}{0,0.08,0.45}

\def\boxit#1{\vbox{\hrule\hbox{\vrule\kern6pt
          \vbox{\kern6pt#1\kern6pt}\kern6pt\vrule}\hrule}}

\definecolor{orange}{rgb}{1,0.5,0}
\definecolor{MyDarkBlue}{rgb}{0,0.08,0.45}

\begin{document}
\title{\Large \bfseries Is the Difference between Deep Hedging and Delta Hedging a Statistical Arbitrage?\thanks{Gauthier is supported by the Natural Sciences and Engineering Research Council of Canada (NSERC, RGPIN-2019-04029), a professorship funded by HEC Montr\'eal, and the HEC Montr\'eal Foundation. Godin is funded by NSERC (RGPIN-2024-04593).}
 } 

\author[a]{Pascal Fran\c cois}
\author[b]{Genevi\`eve Gauthier}
\author[c,d]{Fr\'ed\'eric Godin} 
\author[c]{Carlos Octavio P\'erez Mendoza\thanks{Corresponding author. \vspace{0.2em} \newline
{\mbox{\hspace{0.47cm}} \it Email addresses:} 
\href{mailto:pascal.francois@hec.ca}{pascal.francois@hec.ca} (Pascal François),
\href{mailto:genevieve.gauthier@hec.ca}{genevieve.gauthier@hec.ca} (Geneviève Gauthier), \href{mailto:frederic.godin@concordia.ca}{frederic.godin@concordia.ca} (Fr\'ed\'eric Godin), \href{mailto:carlos.octavio.perez92@gmail.com}{carlos.octavio.perez92@gmail.com} (Carlos Octavio P\'erez Mendoza).}} 


\affil[a]{{\small Research Fellow, Canadian Derivatives Institute and Department of Finance, HEC Montr\'eal, Canada}}
\affil[b]{{\small GERAD and Department of Decision Sciences, HEC Montr\'eal,  Canada}}
\affil[c]{{\small Concordia University, Department of Mathematics and Statistics,  Canada}}
\affil[d]{{\small Quantact Laboratory, Centre de Recherches Math\'ematiques,  Canada}}

\vspace{-10pt}
\date{ 
\today}


\maketitle \thispagestyle{empty} 

%

\vspace{-15pt}

\begin{abstract}
\vspace{-8pt}

{\footnotesize The recent work of \cite{horikawa2024relationship} claims that under a complete market admitting statistical arbitrage, the difference between the hedging position provided by deep hedging and that of the replicating portfolio is a statistical arbitrage. This raises concerns as it entails that deep hedging can include a speculative component aimed simply at exploiting the structure of the risk measure guiding the hedging optimisation problem. We test whether such finding remains true in a GARCH-based market model, which is an illustrative case departing from complete market dynamics. 
We observe that the difference between deep hedging and delta hedging 
is a speculative overlay if the risk measure considered does not put sufficient relative weight on adverse outcomes. Nevertheless, a suitable choice of risk measure can prevent the deep hedging agent from engaging in speculation.}




\noindent \textbf{JEL classification:} C45, C61, G32.


\noindent \textbf{Keywords:} Deep reinforcement learning, optimal hedging, arbitrage.
\end{abstract}

\medskip

\thispagestyle{empty} \vfill \pagebreak

\setcounter{page}{1}
\pagenumbering{roman}

\doublespacing

\setcounter{page}{1}
\pagenumbering{arabic}


\section{Introduction}\label{se:intro}

The seminal paper of \cite{buehler2019deep}, which proposes to use deep reinforcement learning (RL) methods to obtain optimal hedging procedures for financial derivatives, initiated a recent strand of literature.\footnote{See for instance \cite{halperin2019qlbs}, \cite{cao2020deep}, \cite{du2020deep}, \cite{carbonneau2021equal}, \cite{carbonneau2021deep}, \cite{horvath2021deep}, \cite{imaki2021no}, \cite{lutkebohmert2022robust}, \cite{cao2023gamma}, \cite{carbonneau2023deep}, \cite{marzban2023deep}, \cite{mikkila2023empirical}, \cite{raj2023quantum} and \cite{wu2023robust}. 
See also \cite{hambly2023recent} and \cite{pickard2023deep} for related surveys.} Deep RL methods are particulary well-suited to solve dynamic hedging problems because these methods can handle the curse of dimensionality, a problem that more traditional approaches (e.g., finite elements dynamic programming) struggle to overcome. They can also work with very general dynamics for asset prices, not being limited by mathematical tractability issues.

While the ability of deep hedging strategies to outperform standard counterparts is well-documented, the existing literature has not yet extensively analyzed the structure of optimal policies and explained how such incremental performance is attained. \cite{neagu2024deep} make a step in that direction by investigating the impact of the various features on optimal risk management decisions in the presence of illiquidity market impacts. 


In their recent work, \cite{horikawa2024relationship} investigate complete markets that allow for statistical arbitrage with respect to a specific risk measure $\rho$. They assert that, within this framework, deep hedging strategies that minimize the chosen risk metric combine the traditional delta-hedging approach with a statistical arbitrage overlay. In a vector auto-regressive stochastic volatility model and in a GAN-simulated market model, \cite{buehler2021deep} find that the optimal hedging strategy maximizing the entropy utility can also incorporate a statistical arbitrage component.
Such claims raise concerns about the suitability of the deep hedging approach; incorporating speculative or arbitrage-like components that do not contribute to reducing the risk exposure within hedging portfolios would be deemed undesirable in practice. Our objective is therefore to assess empirically whether deep hedging policies minimizing conventional risk metrics still contain a speculative component in incomplete market settings, which would generalize the complete market conclusion of \cite{horikawa2024relationship}. We use a GARCH-based market setting as an illustrative example.


The paper is divided as follows. Section \ref{se:market_model} provides the hedging problem formulation. Section \ref{se:hedging_strategies} describes the deep hedging framework used to solve the problem, and discusses the delta hedging benchmark. Numerical experiments assessing the behavior of the deep versus delta hedging difference strategy are provided in Section \ref{se:SSG}.\footnote{The Python code which allows replicating the numerical experiments from this paper can be found at \href{https://github.com/cpmendoza/DeepHedging_StatisticalArbitrage.git}{https://github.com/cpmendoza/DeepHedging\_StatisticalArbitrage.git}.} Section \ref{se:conclusion} concludes.

\section{Market model for hedging}\label{se:market_model}

This paper considers dynamic risk management strategies for European call options, which involve the construction of a self-financing portfolio composed of the underlying asset and a cash account. The portfolio is rebalanced daily to optimally offset the net risk exposure at the option maturity, denoted as $T$ days. The time-$t$ underlying asset price is $S_t$. The trading strategy is represented by the predictable process 
$\delta=\{\delta_{t}\}_{t=1}^{T}$, where $\delta_{t}$ is the number of underlying asset shares held during the interval $(t-1, t]$. The time-$t$ discounted gain made by the hedging portfolio is $G_{t}^{\delta}=\sum_{k=1}^{t}\delta_{k}(\beta^{k}S_{k}\mbox{e}^{q\Lambda}-\beta^{k-1}S_{k-1})$ with $\beta=\mbox{e}^{-r\Lambda}$, where $r$ is the annualized continuously compounded risk-free rate, $q$ is the annualized underlying asset dividend yield, and the period length is $\Lambda=\frac{1}{252}$ years. The time-$t$ self-financing portfolio value is 
\begin{equation}
\label{eq:portfolio_value}
    V_{t}^{\delta}(V_0) = \beta^{-t}(V_{0}+G_{t}^{\delta}),
\end{equation}
where $V_{0}$ is the initial portfolio value that we set to the option price.

The hedging problem is a sequential decision problem where the holder of a short position in a call option seeks for the best sequence of actions $\delta$ that minimizes the risk associated with the hedging error 
\begin{equation}\label{eq:position}
    \xi_{T}^{\delta} = \mbox{max}(S_{T}-K,0)-V_{T}^{\delta}(V_0),
\end{equation}
where $K$ is the call option strike price. The hedging problem is formulated as
\begin{equation}\label{Hedging_problem}
\delta^{*}=\mathop{\arg \min}\limits_{\delta} \rho  \left(\xi_{T}^{\delta} \right),
\end{equation}
where $\rho$ is the risk measure used by the agent to quantify risk.
In this paper we consider the Conditional Value-at-Risk (CVaR$_\alpha$) defined as $\rho(\xi_{T}^{\delta})=\mathbb{E}[\xi_{T}^{\delta}\mid \xi_{T}^{\delta}\geq \mbox{VaR}_{\alpha}(\xi_{T}^{\delta})]$, where $\alpha \in (0,1)$  and $\mbox{VaR}_{\alpha}(\xi_{T}^{\delta})$ is the Value-at-Risk defined as $\mbox{VaR}_{\alpha}(\xi_{T}^{\delta}) = \min_{c}\{ c:\mathbb{P}(\xi_{T}^{\delta}\leq c)\geq \alpha \}$. The CVaR is a commonly used objective function in the deep hedging literature, see for instance \cite{carbonneau2021equal}, \cite{cao2023gamma} or \cite{wu2023robust}. In addition, an appealing feature of the CVaR is that it allows to finetune the investor’s attitude towards risk through the confidence level. A high value of $\alpha$ puts more emphasis on risk reduction, whereas a low value of $\alpha$ penalizes losses and rewards gains.

Each time-$t$ action $\delta_{t+1}$ is a feedback-type decision, being a function of the information currently available on the market: $\delta_{t+1}= \tilde\delta(X_t)$ for some function $\tilde\delta$ of the state variable vector $X_t$.

\section{Hedging strategies}\label{se:hedging_strategies}

\subsection{Deep hedging}\label{subsec:deep_hedging}

The deep hedging (DH) framework, introduced by \cite{buehler2019deep}, provides a solution to the hedging problem \eqref{Hedging_problem} by leveraging RL techniques. 
The DH policy $\tilde\delta$ is approximated with a neural network $\delta^{DH}_\theta$ with parameters $\theta$, which returns a hedging position $\delta_{t+1}$ when provided with time-$t$ input features $X_{t}$. The objective function to be minimized is thus
\begin{equation}\label{penaltyfunction}
    \mathcal{O}(\theta) = \rho \left(\xi_{T}^{\delta^{DH}_{\theta}} \right).
\end{equation}
The neural network is optimized with the Mini-batch Stochastic Gradient Descent method (MSGD). This training procedure relies on updating iteratively all the trainable parameters of the network  based on the recursive equation
\begin{equation}\label{updatingrule}
    \theta_{j+1} = \theta_{j}-\eta_{j}\nabla_{\theta} \hat{\mathcal{O}}(\theta_{j}),
\end{equation}
where $\theta_{j}$ is the set of parameters obtained after iteration $j$, $\eta_{j}$ is the learning rate (step size) which determines the magnitude of change in parameters on each time step, 
$\nabla_{\theta}$ is the gradient operator with respect to $\theta$ and $\hat{\mathcal{O}}$ is the Monte Carlo estimate of the objective function \eqref{penaltyfunction} computed on a mini-batch. Automatic differentiation packages are used to compute the gradient of $\hat{\mathcal{O}}$. Additional details are provided in the appendix.

For the neural network, we employ a fully-connected Feedforward Neural Network (FFNN) architecture with four hidden layers of width 56 using a ReLU activation function. 
The output FFNN layer, which maps the output of the hidden layer $Z$ into the position in the underlying asset position $\delta_{t+1}^{DH}$, is equipped with a dynamic upper bound on the activation function to preclude excessive leverage. 
Indeed, agents have finite borrowing capacity in practice. We impose that the time-$t$ cash account value $\phi_{t}$ satisfies $\phi_{t}\geq -B$ for all $t$ and for some threshold $B>0$. This is achieved by setting the final output layer activation to
\begin{equation}
   f(Z,t)=\mbox{min}\left(Z,(V_{t}+B)/S_{t}\right), 
\end{equation}
which ensures that the cash amount borrowed in the portfolio remains below $B>0$ (see \cite{François2025}).

Agents are trained on training sets of 400,000 independent simulated paths with mini-batch size of 1,000 and a learning rate of 0.0005 that is progressively adapted with the ADAM \citep{kingma2014adam} optimization algorithm.
The training procedure is implemented in Python, using Tensorflow and considering the \cite{glorot2010understanding} random initialization of the initial parameters of the neural network. Numerical results are obtained from test sets of 100,000 independent paths. 


\subsection{Delta hedging}\label{subsec:delta_hedging}

Delta hedging aims to reduce the risk associated with price movements of an underlying asset by adjusting the hedging portfolio positions in the underlying asset based on the sensitivity ($\Delta$) of the option price to changes in the price of the underlying asset.  Specifically, the time-$t$ position in the underlying asset is defined as the time-$t$ sensitivity $\Delta_{t}$, which is the partial derivative of the time-$t$ option price with respect to the underlying asset value. 

\subsection{Statistical arbitrage}\label{subsec:statistical_arbitrage}

Statistical arbitrage strategies, also known as "good deals" according to the terminology of \cite{cochrane2000beyond}, are profit-seeking trading strategies that capitalize on statistical anomalies in the market. \cite{bondarenko2003statistical} defines statistical arbitrage as a trading strategy that makes a profit on average without requiring any initial capital investment. \cite{assa2013hedging} extend this definition to offer a more nuanced and comprehensive evaluation, ensuring that the trading strategy is not only profitable on average, but also resilient in terms of risk management. As in \cite{assa2013hedging}, we say that $\delta$ is a statistical arbitrage opportunity if
\begin{equation}
    \rho\left(-V_{T}^{\delta}(0)\right)<0,
\end{equation}
that is, if the trading strategy $\delta$ which requires zero initial investment is deemed strictly less risky than a null investment according to risk measure $\rho$.
Such definition is also in line with that of \cite{buehler2021deep}, who focus on the case of the entropy risk measure.

\cite{horikawa2024relationship} claim that in a complete market model that admits statistical arbitrage, the difference between the deep hedging and the delta hedging strategies denoted by
\begin{equation}\label{eq:statistical_arbitrage}
    \delta^{-} = \delta^{DH} - \Delta,
\end{equation}
is a statistical arbitrage strategy according to risk measure $\rho$. We wish to further extend their study and examine if the trading strategy $\delta^{-}$ behaves like a statistical arbitrage in more general incomplete market dynamics, using a GARCH-based market model as a representative candidate for illustration. In other words, we investigate whether the deep hedging approach typically incorporates a speculative arbitrage-like component aimed at exploiting the structure of the risk measure considered. 


\section{Numerical study}\label{se:SSG}

\subsection{Stochastic market dynamics}\label{subse:market_dynamics}

We consider market dynamics based on a GARCH process to represent the underlying asset log-returns. 
The GJR-GARCH(1,1) model introduced by \cite{glosten1993relation} captures time-varying volatility and accounts for the leverage effect. For $t=1,\dots,T$, log-returns in the model follow
\begin{equation}
     R_{t} =\mu + \sigma_{t}\epsilon_{t},\quad \sigma_{t+1}^{2} = \omega + \sigma_{t}^{2}(\alpha + \gamma \mathds{1}_{\{\epsilon_{t}<0\}})\epsilon_{t}^{2}+\beta\sigma_{t}^{2},
\end{equation}
where $\mu,\, \gamma \in \mathbb{R}$, $\omega,\, \alpha,\, \beta$ are positive, $\mathds{1}_{A}$ is the dummy variable indicating if event $A$ occurs and $\{\epsilon_{t}\}_{t=1}^{T}$ are independent standard normal random variables. 
Parameter estimates are obtained through maximum likelihood on a daily time series of the S\&P 500 index extending from January 4, 2016, to December 31, 2020. 
Estimated parameters are $\mu = 0.06\%$, $\omega = 0.01\%$, $\alpha = 0.11$, $\gamma=0.20$ and $\beta=0.78$. 
Furthermore, in all experiments, the annualized continuously compounded risk-free rate and dividend yield are assumed to be constant with values set to $r=2.66\%$ and $q=1.77\%$, respectively. These values represent the historical averages of the 1-year zero-coupon yield and the annualized S\&P 500 dividend yield over the period extending from 1996 to 2020.

The initial option price is computed using Monte Carlo simulation based on the risk-neutral valuation formula
\begin{equation}\label{eq:grach_call}
    \mbox{Call}_{0}=\mbox{e}^{-rT \Lambda}\mathbb{E}^{\mathbb{Q}}[\mbox{max}(S_{T}-K,0)],
\end{equation}
where $\mathbb{Q}$ is a risk-neutral measure.\footnote{The $\mathbb{Q}$ risk-neutral dynamics of the GARCH model are defined by the following equations:
\begin{equation*}
    R_{t} =(r-q)\Lambda - \frac{\sigma^{2}_t}{2}+ \sigma_{t}\tilde{\epsilon}_{t},\quad \sigma_{t+1}^{2} = \omega + \sigma_{t}^{2}(\alpha + \gamma \mathds{1}_{\{\tilde{\epsilon_{t}}-\eta_{t}<0\}})(\tilde{\epsilon_{t}}-\eta_{t})^{2}+\beta\sigma_{t}^{2},
\end{equation*}
where $\eta_{t}=(\mu-(r-q)\Lambda+\sigma^{2}_t/2)/\sigma_{t}$ and $\{\tilde{\epsilon_{t}}\}_{t=1}^{T}$ are independent standard normal random variables under $\mathbb{Q}$. 
}

The call option delta is also calculated through Monte-Carlo simulation based on the relationship
\begin{equation}\label{eq:grach_delta}
    \Delta_{t}=\mbox{e}^{-r\tau \Lambda}\mathbb{E}^{\mathbb{Q}}\left[\frac{S_{t+\tau}}{S_{t}}\mathds{1}_{\{S_{t+\tau}>K\}}\mid \mathcal{F}_{t} \right]
\end{equation}
where $\tau=T-t$ and $\mathcal{F}_{t}$ represents the available information at time $t$, i.e., that being generated by the state $X_t$.

In this model the state space considered for the RL approach is represented by the vector $X_{t}=(V_{t}^{\delta},\, \mbox{log}(S_{t}),\, \sigma_{t+1},\,  \tau)$.


\subsection{Comparative analysis of deep hedging and delta hedging strategies}

In this section, we study the relationship between delta hedging and deep hedging. More specifically, we investigate whether the difference between both strategies represents a speculative overlay reminiscent of a statistical arbitrage. 
The comparison is conducted by hedging the at-the-money (ATM) European call option, with $S_0=K=100$, and maturity $T= 63$ days. 
The leverage constraint is $B = 100$.

\autoref{table:arbitrage} presents the hedging performance of the deep hedging agents trained with the CVaR$_\alpha$ risk measure. We consider the following confidence levels $\alpha$: 1\%, 5\%, 10\%, 20\%, 50\%, 85\%, 90\%, and 95\%. High values of $\alpha$ only put weight on the most adverse outcomes and entail focusing purely on risk reduction. Conversely, low values for $\alpha$ both penalize losses and reward gains, which leads to seeking risk-reward trade-offs. As such, the CVaR$_\alpha$ with a low confidence level does not align with the objective of limiting the variability of the hedging error. Since the CVaR$_\alpha$ is an increasing function of $\alpha$, there are more statistical arbitrage strategies becoming available as $\alpha$ decreases.

Columns labeled "Base strategies" display the risk measure applied to the hedging error for the deep hedging strategy, and the difference between the risk provided by deep hedging and that of delta hedging. Columns labeled "Difference strategy" provide statistics (hedging error risk and expectation of net cash flow) of the trading strategy $\delta^{-}$ representing the differential position between deep hedging and delta hedging. Since such strategy reflects a long position on the deep hedge and a short position on the delta hedge, the option payoffs from the two (long and short) positions cancel out. Hence, performance is assessed by looking at $V_{T}^{\delta^{-}}(0)$ rather than $\xi_{T}^{\delta^{-}}$.


\begin{table}[H]
\centering
\renewcommand{\arraystretch}{1.5}
\caption{Performance assessment for deep hedging, delta hedging and their difference over a short position on an ATM call option with maturity $T=63$ days.}
\label{table:arbitrage}
\begin{tabular}{{p{1.7cm}>{\centering\arraybackslash}p{2.7cm}>{\centering\arraybackslash}p{2.7cm}c>{\centering\arraybackslash}p{2.7cm}>{\centering\arraybackslash}p{2.7cm}}}

\hline
 \multicolumn{1}{c}{} & \multicolumn{2}{c}{Base strategies} & & \multicolumn{2}{c}{Difference strategy} \\

\cline{2-3}\cline{5-6}

Metric & $\rho(\xi_{T}^{\delta^{DH}})$ & $\rho(\xi_{T}^{\delta^{DH}})-\rho(\xi_{T}^{\Delta})$ & & $\rho(-V_{T}^{\delta^{-}}(0))$ & $\mathbb{E}[V_{T}^{\delta^{-}}(0)]$ \\
\hline
CVaR$_{1\%}$ & -1.918 & -1.368 &   & -1.306 & \phantom{-}1.372 \\ 
CVaR$_{5\%}$ & -1.852 & -1.351 &   & -1.167 & \phantom{-}1.372 \\ 
CVaR$_{10\%}$ & -1.767 & -1.325 &   & -1.038 & \phantom{-}1.371 \\ 
CVaR$_{20\%}$ & -1.575 & -1.256 &   & -0.807 & \phantom{-}1.371 \\ 
CVaR$_{50\%}$ & -0.614 & -0.791 &   & \phantom{-}0.071 & \phantom{-}1.370 \\ 
CVaR$_{85\%}$ & \phantom{-}1.505 & -0.129 &   & \phantom{-}1.208 & -0.039 \\ 
CVaR$_{90\%}$ & \phantom{-}2.055 & -0.128 &   & \phantom{-}1.599 & -0.225 \\ 
CVaR$_{95\%}$ & \phantom{-}3.102 & -0.121 &   & \phantom{-}2.121 & -0.369 \\ 
\hline
\end{tabular}
\begin{tablenotes}
\item 
Results are computed using 100,000 out-of-sample paths. The initial price of the option is 3.16. 
$\xi_{T}^{\delta}$ is the hedging error for trading strategy $\delta$, with $\delta^{DH}$ being deep hedging and $\Delta$ being delta hedging. The strategy $\delta^{-}$ uses the underlying asset position defined by the difference between that of the deep hedging and the delta hedging strategies. 
\end{tablenotes}
\end{table}

For all confidence levels $\alpha$ below $50\%$, the strategy $\delta^{-}$ exhibits both positive average profitability $\mathbb{E}[V_{T}^{\delta^{-}}(0)]$ and a CVaR value $\rho(-V_{T}^{\delta^{-}}(0))$ that is negative. This corresponds to a formal statistical arbitrage strategy. Moreover, for the case $\alpha =50\%$, even if the risk measure is positive, it is nevertheless negligible in comparison to expected profits. The strategy $\delta^{-}$, though not a statistical arbitrage from the definition, exhibits a behavior that is quite similar to a statistical arbitrage. This might indicate that the \cite{buehler2021deep} approach, which consists in using a change of measure under which statistical arbitrage strategies are removed, might be insufficient to prevent speculative behavior from the hedging agent; such speculative strategies do not always qualify as formal statistical arbitrages.
Conversely, difference strategies $\delta^{-}$ using $\alpha \geq 85\%$ clearly do not qualify as statistical arbitrage; the associated risk measure is high and the average profitability is negative. 

\begin{figure}[H]\centering
\caption{P\&L distribution of the strategy $\delta^{-}$.}
\label{fig:p&l}
\includegraphics[width=17.3cm]{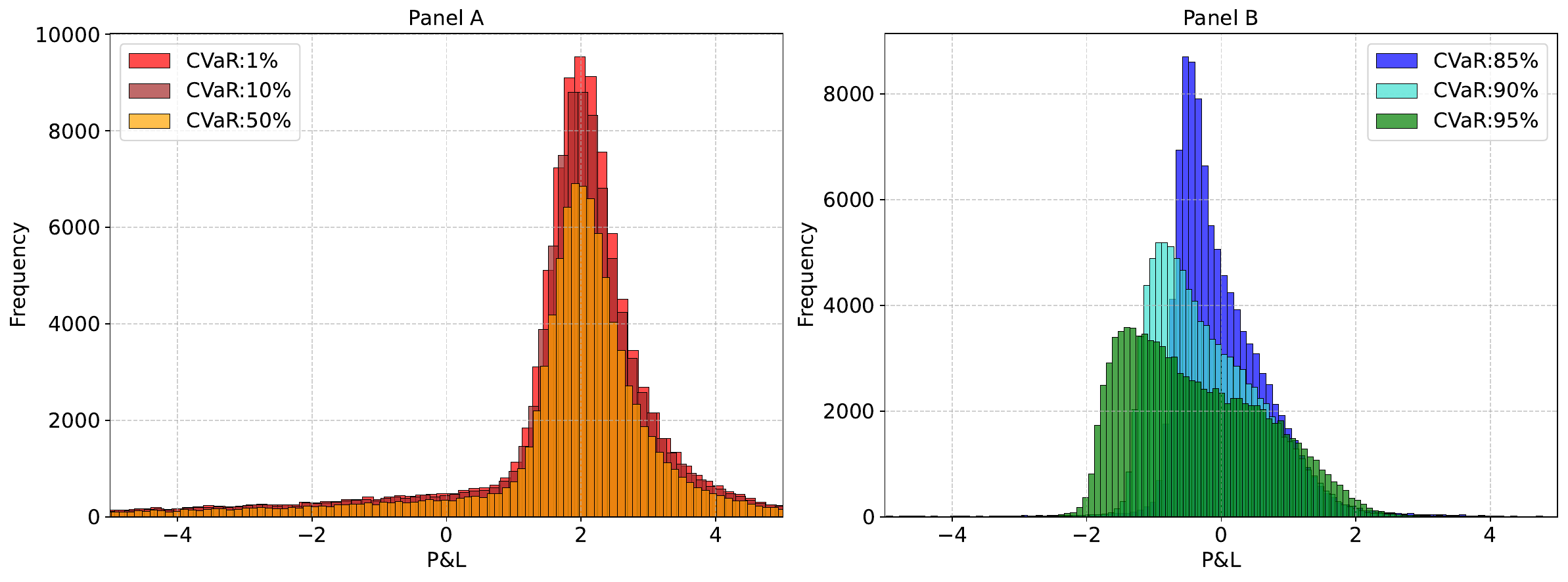}
\begin{tablenotes}
\item Distributions are computed using 100,000 out-of-sample paths. 
The P\&L is simply defined by the portfolio value $V_{T}^{\delta^{-}}(0)$ at maturity.
\end{tablenotes}
\end{figure}

The distribution of the profit and loss (P\&L) for the trading strategy $\delta^{-}$, which is $V_{T}^{\delta^{-}}(0)$, is depicted in \autoref{fig:p&l} for various confidence levels. This confirms that difference strategies associated with low confidence levels $\alpha$ (1\%, 10\% and 50\%) are exactly or similar to statistical arbitrage with very high average profits and a very fat left tail (high extreme loss potential). 
The deep hedging agent is incorporating a strong speculative element in its trading strategy, which is unsuitable in practice. 
Conversely, the strategies associated with higher values for $\alpha$ do not exhibit characteristics of a statistical arbitrage and do not lead to concerns about the suitability of the deep hedging strategy.

We analyze the statistical relationship between deep hedging and delta hedging strategies 
through (i) sample Spearman (rank) correlations between underlying asset positions of both strategies, and (ii) the regression model
\begin{equation}
    \delta^{DH} = \kappa_{0}+\kappa_{1}\Delta+\epsilon,
\end{equation}
with $\delta^{DH}$ and $\Delta$ being positions produced by the deep hedging and delta hedging strategies, respectively.
Metrics (regressions and correlations) are computed across all rebalancing points of all paths in the test sets.

\autoref{table:aggregated_metrics} provides the Spearman correlation coefficient $\varrho$, which evaluates monotonic relationships between strategies, and the coefficient of determination $R^{2}$, which measures the strength of the linear association between the strategies. 
These metrics are computed for the various CVaR confidence levels.

\begin{table}[H]
\centering
\renewcommand{\arraystretch}{1.5}
\caption{Statistical relationships between positions of delta hedging and deep hedging.}
\label{table:aggregated_metrics}
\begin{tabular}{{p{1.5cm}>{\centering\arraybackslash}p{2.5cm}>{\centering\arraybackslash}p{2.5cm}}}

\hline
 \multicolumn{1}{c}{} & \multicolumn{2}{c}{Statistics} \\

\cline{2-3}

Metric & $\varrho$ & $R^{2}$  \\
\hline
CVaR$_{1\%}$ & -0.270 & 0.003  \\ 
CVaR$_{5\%}$ & -0.271 & 0.003  \\ 
CVaR$_{10\%}$ & -0.272 & 0.003 \\ 
CVaR$_{20\%}$ & -0.273 & 0.003 \\ 
CVaR$_{50\%}$ & -0.273 & 0.003 \\ 
CVaR$_{85\%}$ & \phantom{-}0.939 & 0.773  \\ 
CVaR$_{90\%}$ & \phantom{-}0.963 & 0.816  \\ 
CVaR$_{95\%}$ & \phantom{-}0.969 & 0.808  \\ 

\hline
\end{tabular}
\begin{tablenotes}
\item 
Results are for a short position on the ATM call option with a maturity of $N=63$ days. They are computed using 100,000 out-of-sample paths. The metric $\varrho$ denotes the (unconditional) Spearman correlation between underlying asset positions of the delta hedging strategy and the deep hedging strategy across all rebalancing days while $R^{2}$ represents the 
$R^{2}$ statistic obtained after regressing deep hedging positions onto delta hedging positions.
\end{tablenotes}
\end{table}
Results presented in \autoref{table:aggregated_metrics} show strong monotonic and linear association between deep and delta hedging positions for high confidence levels $\alpha=85\%$, $90\%$ or $95\%$. The deep hedging strategies can therefore be seen as alterations of the delta hedging procedure that improve hedging performance. Conversely, for low confidence levels ($50\%$ or below), deep hedging positions seem completely unrelated to delta hedging positions, indicating that the agent has mostly abandoned its hedging objective and is rather attempting to speculate or conduct statistical arbitrage.

\section{Conclusion}\label{se:conclusion}

Consider the trading strategy whose underlying asset positions correspond to the difference between these of deep hedging and delta hedging. What if there exist market models
under which such strategy is a statistical arbitrage? This would raise concerns about the suitability of deep hedging procedures, as it raises the possibility that typical deep hedging strategies could consist of conventional hedging strategies that are enhanced with speculative overlays which are unrelated to hedging. 

Our study shows that these concerns can be mitigated under GARCH-based market models; if the risk measure considered in the hedging optimization problem does not sufficiently penalize losses relative to rewards provided for gains, the deep hedging strategy attaches a statistical arbitrage strategy overlay to the delta hedging strategy.
Nevertheless, when using a proper risk measure (the CVaR with sufficiently high $\alpha$ in our case) within the optimization problem, the difference between deep hedging and delta hedging does not exhibit statistical arbitrage-like behavior and cannot be interpreted as a speculative strategy reaping profits while exploiting blind spots of the chosen risk measure. 

The two main conclusions from this study are therefore that (i) the objective function of the deep hedging problem must be carefully selected to prevent the hedging agent from abandoning its hedging objective and pursuing speculative behavior, and (ii) deep hedging can soundly achieve its hedging objectives when provided with a suitable risk measure. A possibility could be to use risk measures that do not provide any reward for gains, such as the semi-RMSE used in \cite{carbonneau2023deep}. However, this would come with the cost of negatively impacting the profitability of the strategy. More research is therefore required to determine what risk measure could be used in the objective function to produce sound hedging behavior.


\bibliographystyle{apalike}
\bibliography{references}  


\appendix

\section*{Appendix: Details for the MSGD training approach}\label{appen:MSGDtraining}

The MSGD method estimates the penalty function $\mathcal{O}(\theta)$, which is typically unknown, through small samples of the hedging error called batches. Let $\mathbb{B}_{j}=\{ \xi_{T,i}^{\delta^{DH}_{\theta_{j}}} \}_{i=1}^{N_{\mbox{\scriptsize{batch}}}}$ be the $j$-th batch where $\xi_{T,i}^{\delta^{DH}_{\theta_{j}}}$ denotes the hedging error of the $i$-th simulated path in the $j$-th batch defined as
$$\xi_{T,i}^{\delta^{DH}_{\theta_{j}}} = \mbox{max}(S_{T,ij}-K,0)- V_{T,i}^{\delta^{DH}_{\theta_{j}}}(V_0)\mbox{,}$$
where $S_{T,ij}$ is the price of the underlying asset at time $T$ in the $i$-th simulated path, and $V_{T,i}^{\delta^{DH}_{\theta_{j}}}$ is the terminal value of the hedging strategy for that path when $\theta = \theta_{j}$. The penalty function estimation for the batch $\mathbb{B}$ is
\begin{equation*}
    \Hat{C}^{\left(\mbox{\scriptsize{CVaR}}\right)}(\theta_{j},\mathbb{B}_{j})= \widehat{\mbox{VaR}}_{\alpha}(\mathbb{B}_{j})+\frac{1}{(1-\alpha)N_{\mbox{\scriptsize{batch}}}}\sum_{i=1}^{N_{\mbox{\scriptsize{batch}}}}\max\left( \xi_{T,i}^{\delta^{DH}_{\theta_{j}}}-\widehat{\mbox{VaR}}_{\alpha}(\mathbb{B}_{j}),0 \right)\, \mbox{,}
\end{equation*}

where $\widehat{\mbox{VaR}}_{\alpha}(\mathbb{B}_{j})=\xi_{T,\lceil \alpha \cdot N_{\mbox{\scriptsize{batch}}} \rceil}^{\delta^{DH}_{\theta_{j}}}$ is the estimation of the VaR obtained from the ordered sample $\{ \xi_{T,[i]}^{\delta^{DH}_{\theta_{j}}} \}_{i=1}^{N_{\mbox{\scriptsize{batch}}}}$ and $\lceil \cdot \rceil$ is the ceiling function. These empirical approximations are used to estimate the gradient of the penalty function required in Equation \eqref{updatingrule}.\footnote{Details about gradient of the empirical objective function are provided in \cite{goodfellow2016deep}.} 






\end{document}